\documentclass[aps,reprint,showpacs,amsmath,amssymb,twocolumn]{revtex4-1}
\usepackage[utf8x]{inputenc}
\usepackage{graphicx}
\usepackage{amsmath}
\usepackage{dcolumn}
\usepackage{bm}

\begin{document}

\title{Vortices in Bose-Einstein condensates - finite-size effects and the thermodynamic limit} 

\author{J.C. Cremon$^1$, G.M. Kavoulakis$^2$, B.R. Mottelson$^3$ and
  S.M. Reimann$^1$}
\affiliation{$^1$Mathematical Physics, Lund Institute of Technology, 
P.O. Box 118, SE-22100 Lund, Sweden \\
$^2$Technological Education Institute of Crete, P.O.
Box 1939, GR-71004, Heraklion,
Greece \\
$^3$The Niels Bohr International Academy, Blegdamsvej 17, DK-2100 Copenhagen \O, Denmark}
\date{\today}

\begin{abstract}
For a weakly-interacting Bose gas rotating in a harmonic trap 
we relate the yrast states of small systems (that can be treated exactly) 
to the thermodynamic limit (derived within the mean-field approximation). 
For a few dozens of atoms, the yrast line shows distinct quasi-periodic 
oscillations
with increasing angular momentum that originate from the internal structure of 
the exact many-body states. These finite-size effects disappear in the
thermodynamic limit, where the Gross-Pitaevskii approximation provides the 
exact energy to leading order in the number of particles $N$. 
However, the exact yrast states reveal significant structure not 
captured by the mean-field approximation: Even in the limit of large $N$,
the corresponding mean-field solution accounts for  
only a fraction of the total weight of the exact quantum state. 
\end{abstract}


\maketitle

\section{Introduction}
Contrary to many other systems with superfluid properties like, e.g., 
liquid Helium or atomic nuclei, ultra-cold atomic quantum gases are 
-- at least under typical conditions -- very dilute.  
Still, they may exhibit superfluid properties \cite{leggett1999,leggett2001}
because of their ultralow temperatures.
Initial experiments with trapped Bose-Einstein 
condensates~\cite{davis1995a,davis1995b,andersson1995,cornellwieman2002,ketterle2002}
have been performed mainly in large systems confining 
thousands to millions of atoms.
It was only more recently that experiments reached  
the limit of smaller atom numbers $N \sim {\cal O} (1)$~\cite{jochim2011}. 
In small systems, however, the thermodynamic limit often 
applied to the case of homogeneous superfluids is not 
appropriate. Even in the regime of weak interactions, 
deviations from this limit are expected due to finite-size
effects and the influence of the trapping potential. 

The rotational properties of Bose-Einstein condensates in a harmonic trap 
have been studied extensively in the past, see the
reviews~\cite{fetter2009,bloch2008,cooper2008,viefers2008,saarikoski2010}. 
Previous theoretical studies applied the Gross-Pitaevskii method
(for example,~\cite{butts1999,kavoulakis2000,linn1999,linn2001,garcia2001}), 
or have gone beyond the mean-field approximation
(for example,~\cite{wilkin1998,mottelson1999,bertsch1999,jackson2000,smith2000,cooper2001,jackson2001,manninen2001,ueda2001,liu2001,nakajima2001,nakajima2003,romanovsky2004,manninen2005,reimann2006a,reimann2006b,barberan2006,romanovsky2006,ueda2006,yannouleas2007,dagnino2007,hamamoto2007,romanovsky2008,parke2008,dagnino2009a,dagnino2009b,liu2009,nunnenkamp2010,baym2010,baym2013}). 
Most of these studies made use of 
the numerical ``exact'' diagonalization of the many-body 
Hamiltonian, which we also employ here.  
Reference~\cite{baym2010} and the more recent study in~\cite{baym2013} 
examined the effect of correlations on the rotational properties  
considering Bogoliubov fluctuations on the mean-field state. 
Reference~\cite{kwasigroch2012} went beyond the Bogoliubov
description and considered interactions between the quasi-particle
excitations.

As discussed below, the exact quantum states of a few dozens of weakly 
interacting atoms in a rotating harmonic trap reveal 
significant structure not captured by the Gross-Pitaevskii
approxi\-mation. 
It is well known from earlier studies of mean-field theory in the thermodynamic
limit that with increasing rotational frequency, a dilute Bose gas 
in a harmonic trap goes through a systematic series of phase 
transitions associated with the formation of 
vortices~\cite{butts1999,kavoulakis2000,linn1999,linn2001,garcia2001}.
In the limit of small $N$, however, finite-size effects 
become important: Quasi-periodic oscillations occur 
along the ``yrast line'' connecting the lowest-energy states 
as a function of angular momentum
(\cite{liu2001,manninen2005,reimann2006a,reimann2006b,barberan2006,romanovsky2008,liu2009},
see also the discussion in the review 
articles~\cite{cooper2008,saarikoski2010}).
These oscillations lead to discontinuous
steps in the ground state angular momentum $L$ 
as a function of the trap rotation frequency $\Omega $.
They originate from the 
structure of the exact many-body wave function, generalizing the well-known 
pattern first described by Butts and Rokshar in the mean-field 
thermodynamic limit~\cite{butts1999}. 
The Gross-Pitaevskii approximation 
is known to provide the exact energy to leading 
order in $N$~\cite{lieb2000,jackson2001,lieb2009}. 
However, we find that only a fraction of the total weight of the 
exact quantum state accounts for the corresponding mean-field solution  
even in the limit of rather large numbers of atoms. 

\section{Model}

We consider $N$ bosons of mass $M$, confined in a harmonic oscillator potential 
that is isotropic in two dimensions $(x,y)$, with $z$ 
taken to be the axis of rotation of the cloud. 
We assume that the system is quasi two-dimensional, with 
the motion along the $z$-axis being frozen (i.e. oscillator frequencies
$\omega = \omega _x =\omega _y\ll \omega _z$ and $\hbar \omega _z$  
larger than the interaction energy).  
For sufficiently weak interactions one may restrict the set of 
single-particle states of the harmonic potential to those with 
zero radial nodes, which is the so-called lowest-Landau-level 
approximation~\cite{fetter2009}. Then, the quantum number $m\ge 0$
specifying the $z$-component of single-particle 
angular momentum is the only quantum number defining the orbitals 
$\psi _{0,m}\propto r^m \hbox{e}^{im\phi } \hbox{e}^{-r²/{2\ell ^2}}$ 
(with $\ell = \sqrt {\hbar/M\omega }$). 
The set ${\cal F}$ of Fock states
$\{ |\Phi _j \rangle \}_{j=1}^{F} = 
\{ |0^{N_0}, 1^{N_1}, \dots m^{N_m} \rangle \}_{j=1}^{F}$ 
(where $N_m$ denotes the number of particles in a single-particle state with 
angular momentum $m$)  labeled by the index $j$ spans the 
basis of the many-body state. These Fock states 
are chosen as eigenstates of the particle-number operator
and of the angular-momentum operator, with  
$\sum_m N_m = N$ as well as $\sum_m m N_m = L$ (units of $\hbar $). 
In the absence of interactions there is a large degeneracy 
which comes from the different ways that one may distribute 
$L$ units of angular 
momentum to $N$ particles in a harmonic potential~\cite{mottelson1999}. 
Clearly, this degeneracy increases with increasing $L$ and $N$. 
In the spirit of degenerate perturbation theory, 
the Hamiltonian ${\hat H}_{\mathrm{rot}}$ is diagonalized in the 
subspace of these  
degenerate states. For effective contact interactions 
between the bosonic atoms~\cite{lewin2009},  
in the rotating frame of reference,  ${\hat H}_{\mathrm{rot}}$ is given by 
\begin{equation}
{\hat H}_{\mathrm{rot}} = 
\hbar \omega N  + \hbar (\omega -\Omega ){L} 
+ {g\over 2}   \sum _{k\ne l} \delta ({\bf r}_k - {\bf r} _l)  ~,  
\end{equation}  
i.e., only the part coming from the two-body interactions 
needs to be diagonalized.
Here, $\Omega $ is the trap rotation frequency, and 
$g = U_0 \int |\phi_0(z)|^4 \, dz = U_0/(\sqrt{2 \pi} \ell_z)$ is the
interaction strength, with  
$\phi_0(z)=e^{-z^2/2\ell_z^2}/(\pi \ell_z^2)^{1/4}$ being the ground state of the 
potential along the $z$ axis, and $\ell_z$ the 
oscillator length in the $z$ direction. 
Also, $U_0 = 4 \pi \hbar^2a/M$ is the matrix element 
for zero-energy elastic two-body collisions between the 
atoms, with $a$ being the corresponding scattering length. 
We can thus define the dimensionless parameter 
$\lambda = NMg/\hbar^2 = \sqrt{8\pi } N a /l_z$ to measure 
the coupling strength. The eigenstates of ${\hat H}_{\mathrm{rot}}$  
are expressed as 
$|L,N\rangle  = \sum _{j=1}^{F} C_{j} \mid {\Phi _j}\rangle $.
\begin{figure}
\includegraphics[width=0.95\columnwidth]{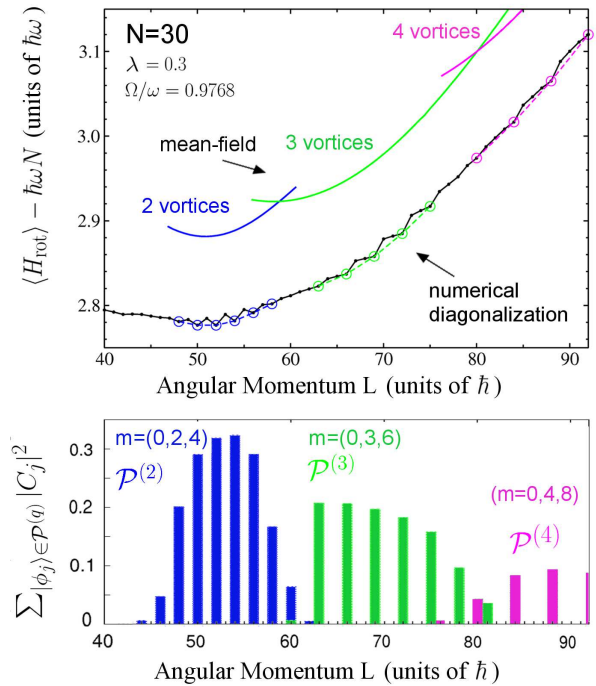}
\caption{(Color online) Comparison of mean-field and ``exact'' 
yrast states of $N=30$. 
{\it Upper panel:} 
Energy (in the rotating frame) as a function of total 
angular momentum $L$,
at a rotational frequency of $\Omega/\omega = 0.9768$ and 
a coupling strength $\lambda = 0.3$. 
The black line shows the result 
of the numerical diagonalization.
Cusps in the yrast line occur with a quasi-periodicity 
of $q=2, 3$ and $4$ in $L$, as marked by the {\it blue}, {\it green} 
and {\it magenta} circles.
The three higher-energy parabolae 
show the result of the corresponding 
mean-field variational calculation.
{\it Lower panel:} 
Sum $\sum_{|\Phi _j \rangle \in {\cal P}^{(q)}} |C_j|^2$ of amplitudes of 
Fock states $|\Phi _j\rangle $ built exclusively out of the orbitals 
that are macroscopically occupied within the 
mean-field approximation (see text), with 
$(m = 0,2,4)$, $(m = 0, 3, 6)$, and $(m=0,4,8)$ 
for the cases of two-fold ({\it blue, $q=2$} ), three-fold ({\it green, $q=3$}) 
and four-fold ({\it magenta, $q=4$}) symmetry, respectively.  
\\
}
\label{fig:1}
\end{figure}
\section{Results}
It is instructive to start with the case $L/N = 1$, where there 
is a single vortex state at the center of the cloud, the so-called 
``unit vortex''. Within the mean-field approximation 
\cite{butts1999,kavoulakis2000},  all the atoms 
reside in one single-particle state with 
$m=1$.
However, the exact many-body wavefunction (which in this case 
is known analytically, see~\cite{wilkin1998,bertsch1999,smith2000}) 
has a different structure.  
Although the dominant Fock state corresponds to the 
mean-field state with macroscopic occupancy of the $m=1$ orbital, in 
addition there are orbitals with $m=0$ and $m=2$.
The yrast state of the unit vortex 
can be written as\cite{wilkin1998,kavoulakis2000}
$| L=N,N \rangle  = \sum_{k} (-1)^k  C_{k} | 0^k, 1^{N-2k}, 2^k \rangle$,
where $C_{k} = {1}/ {(\sqrt{2}^{k+1})}$ to leading order in $N$. 
The component corresponding to the 
mean-field approximation is the single term with $k = 0$, with  
$|C_{0}|^2=1/2$.  All other Fock 
states, with their sum trivially adding up to completeness,
$\sum_{k \neq 0} |C_{k}|^2 = 1/2 $,  
have significantly smaller amplitudes. 
In other words, half of the weight of the total wavefunction
is not captured by the single dominant Fock state carrying a macroscopic 
occupancy that corresponds to the mean-field solution. 
(For a discussion of the unit vortex see also 
Refs.~\cite{bertsch1999,nunnenkamp2010,baym2010}).
Evaluating the mean occupancy of the three single-particle states from the
exact state, one finds \cite{wilkin1998} that to leading order in $N$, 
the occupancy of the $m=1$ state is 
$\langle N_1 \rangle = N-2$, while for $m=0$ and $m=2$ we have 
$\langle N_0 \rangle = \langle N_2 \rangle = 1$.  
(The mean occupancy of all other single-particle states 
is of lower order in $N$, which justifies to neglect them).
Thus, there is only 
one single-particle orbital that is macroscopically occupied for large $N$. 
The depletion of the condensate, defined as $({\langle N_0 \rangle} 
+ {\langle N_2 \rangle})/N$, equals $2/N$.
In the mean-field approximation
the energy (in the laboratory frame of reference) at $L/N=1$ is 
${\cal E}_{\rm MF} = N \hbar \omega + g N (N-1)/2 $~
(see~\cite{kavoulakis2000}), while  
the exact energy is
${\cal E}_{\rm ex} = N \hbar \omega + g N (N-2)/2$ 
(see~\cite{wilkin1998,bertsch1999}).  
The comparison shows that the mean-field energy is correct to leading 
order in $N$, while the contribution of the 
single-particle states $m=0$ and $m=2$ that are absent in the mean-field
solution give corrections to the energy that are of lower
order in $N$~\cite{jackson2001}.  

Beyond the unit vortex, for $L/N > 1$ the yrast states are not
analytically known, and one needs to turn to numerical methods instead. 
The upper panel of Fig.~\ref{fig:1} shows 
the yrast energies (in the rotating frame) 
obtained by the Gross-Pitaevskii 
method (upper, parabolic lines), in comparison to the energies 
obtained by exact diagonalization (lower black line),  
calculated for $N=30$, for the
interaction strength $\lambda = 0.3$, and $\Omega/\omega = 0.9768$
\footnote{Here, up to $m \le 14$ single-particle orbitals were included in the 
basis states. For diagonalization we used the ARPACK library.}.   

For $1.7\stackrel{<}{\sim }L/N\stackrel{<}{\sim }2.03$, 
the yrast state consists of single-particle 
orbitals with even values of $m$, and thus has two-fold symmetry. 
The occupancy of the orbitals with odd $m$ is of lower order in $N$, 
and thus negligible in the thermodynamic limit asssumed within the 
mean-field approximation~\cite{kavoulakis2000}.
For the simple form 
$ \Psi = c_0 \psi_{0,0} + c_2 \psi_{0,2} + c_4 \psi_{0,4}$, 
the mean-field energy is straightforwardly obtained variationally
under the constraints of fixed particle number and of fixed expectation 
value of angular momentum. For two-fold symmetry, the 
corresponding energy is shown as the upper blue line in Fig.~\ref{fig:1}. 
Similarly, we may evaluate the energies for the order parameter
with three-fold symmetry (green line), consisting exclusively of 
single-particle orbitals with angular momenta $m$ that are multiples of three,
$\Psi = c_0 \psi_{0,0} + c_3 \psi_{0,3} + c_6 \psi_{0,6},$ 
and four-fold symmetry (grey line) with $m$ being multiples of four,  
$   \Psi = c_0 \psi_{0,0} + c_4 \psi_{0,4} + c_8 \psi_{0,8}. $ 
The local energy minima associated with a given symmetry 
in the order parameter compete with each other, 
giving rise to the discontinuous phase transitions between 
states of different symmetry as $\Omega $ increases.

For finite $N$, the {\it exact} energy (shown by the black line in
Fig.~\ref{fig:1}) overall lies below the mean-field value, as expected. 
The yrast line shows oscillations with a quasi-periodicity increasing from 
$q=2$ to $3$,  and then $4$ units of angular momentum 
(for the range of $L$ considered here).  
We find that the minima (downward cusps, marked by circles) 
occurring with quasi-periodicity $q$ lie on parabolic energy branches
that are associated with the symmetry of the yrast states,  
similar to the Gross-Pitaevskii mean-field result. 
The crossings between the different branches 
mark the transitions between the different symmetries. 
\begin{figure}
\includegraphics[width=0.95\columnwidth]{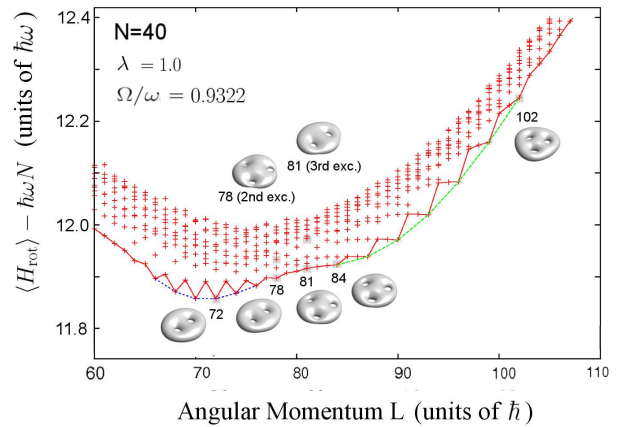}
\caption{(Color online). The energy of the yrast state and of the low-lying 
excited states in the rotating frame 
for $N=40$, $\Omega/\omega = 
0.9322$, and $\lambda = 1.0$. 
Insets: Isosurfaces (placed at half-maximum value) of the pair-correlated
densities; reference point in the $(x,y)$-plane at $(1,0)$. 
(Units of the oscillator length $(\hbar/M \omega)^{1/2}$) (see text).}
\label{fig:2}
\end{figure}

Figure~\ref{fig:2} shows the yrast line and low-lying excitations 
in the rotating frame for $N=40$ particles, for $\lambda = 1.0$, and $\Omega/
\omega=0.9322$ (here for a basis with $m\le 12$), where similar oscillations
occur 
(here only shown up to $q=3$ due to the rapid increase 
in matrix dimension for larger $N$).  
The insets to Fig.~\ref{fig:2} show the isodensity surfaces of the 
pair-correlated densities (defined as 
$\langle \Psi | \hat \Psi ^\dagger ({\bf r})  \hat \Psi ^\dagger ({\bf r}')  
\hat \Psi ({\bf r}')  \hat \Psi  ({\bf r})  |\Psi \rangle $).
Around the transition between two and three vortices
(see insets) it is apparent that there is a crossing of states.  
 
Let us now further analyze the quasi-periodicity of the yrast line 
for the example of the
two-vortex state. If $N$ is even and $L$ is a 
multiple of 2 (but not of 4), then the Fock 
states $|k \rangle$ with the largest amplitudes giving rise to the downward cusps have the form
\begin{eqnarray}
  |0^{k + N/2 - (L+2)/4}, 2^{N+1-2k}, 
  4^{k-N/2+(L-2)/4}
\rangle. 
\label{eq:s1}
\end{eqnarray}
If $L$ is a multiple of 4, then the corresponding states are
\begin{eqnarray}
  |0^{k + N/2 - L/4}, 2^{N-2k}, 4^{k-N/2+L/4}
\rangle.
\label{eq:s2}
\end{eqnarray}
The integer $k$ takes all the possible values for which the 
occupancies are non-negative.

\begin{figure}
\includegraphics[width=0.75\columnwidth]{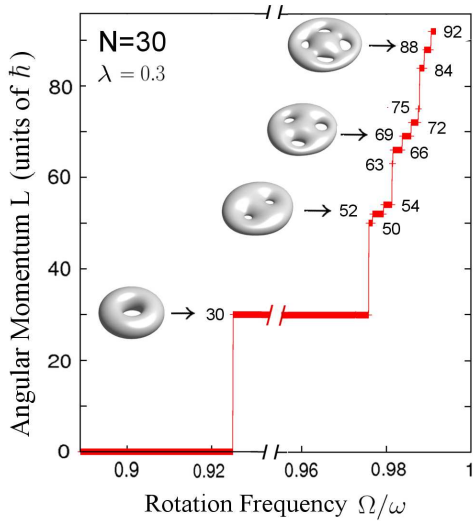}
\caption{(Color online).
Angular momentum $L = L(\Omega/\omega)$  
resulting from minimizing $E_0^{\hbox{rot}}$ for $N = 30$ 
and $\sqrt{8\pi } N a /l_z =0.3$, showing additional steps in $L$ 
that originate from the quasi-periodicity of the yrast line.  
The insets show isosurfaces of the pair-correlated densities (as in Fig.~2).}
\label{fig:3}
\end{figure}

The quasi-periodic oscillations give rise 
to the additional distinct steps (as in this case, 
of two units in $L$ in the region of the 
two-vortex state) in the graph $L(\Omega)$ that is  
obtained by minimizing the energy at a given value of $\Omega $ 
in the rotating frame, see Fig.~\ref{fig:3}. 
A similar situation occurs for the vortex states 
with three- and four-fold symmetry, giving rise to the 
corresponding quasi-periodicity in the yrast energy, 
as well as the steps in $L(\Omega )$.  
These additional steps disappear in the thermodynamic 
limit and the curve becomes a piece-wise continuous 
function of $\Omega$, as described by Butts and Rokshar~\cite{butts1999}. 
Along the steps, the pair-correlated densities (shown as isosurfaces in the
insets to Fig.~\ref{fig:3}) follow a pattern 
similar to the mean field results~\cite{butts1999}. 

\begin{figure}
\includegraphics[width=0.95\columnwidth]{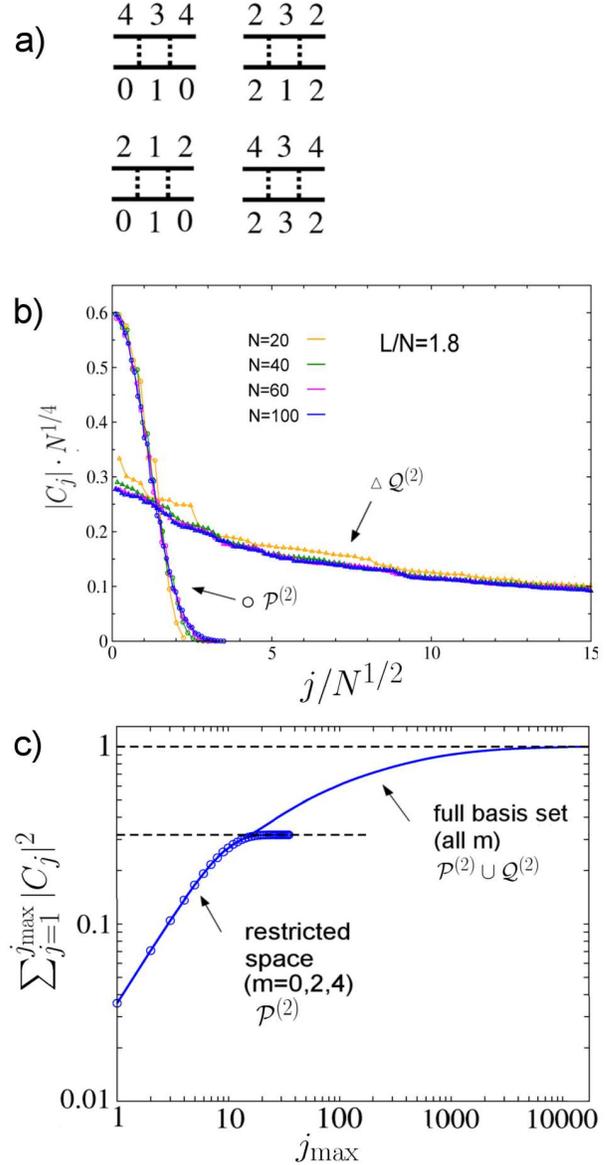}
\caption{(Color online) Comparison between the 
restricted and the complementary Fock space, for
the two-vortex state at $L/N=1.8$. 
(a) Diagrams showing the contributions to subleading order in $N$, here for 
$m\le 4 $. (b)  Amplitudes in the subspaces 
${\cal P}^{(2)}$ and ${\cal Q}^{(2)}$, ordered after their 
absolute size. (c) Saturation of the sum of amplitudes  
$\sum _{j=1}^{j_{max}} |C_j|^2$ (here for $N=100$) 
in ${\cal P}^{(2)}$ to about 30\% of the full weight.}
\label{fig:4}
\end{figure}

In the mean-field solution, for a given symmetry 
only a certain subset of 
single-particle states with angular momentum 
$m$ contribute to the order parameter~\cite{kavoulakis2000}.
The whole Fock space ${\cal F}$ may thus be viewed as composed of a subspace 
${\cal P}^{(q)}$ that is exclusively built on the Fock states constructed with  
single-particle orbitals 
that appear in the mean-field solution 
(as for two-, three- or four-fold symmetry, $q=2, 3$ or $4$, 
these are only the orbitals with $m=(0,2,4)$, 
$m=(0,3,6)$ and $m=(0,4,8)$, respectively),  
and the rest of all the Fock states building a
space that we call ${\cal Q}^{(q)}$. Obviously, ${\cal F} = {\cal P}^{(q)} 
\cup {\cal Q}^{(q)}$. (For the two-vortex case, 
the Fock states in ${\cal P}^{(2)}$ 
were given in Eqs.~(\ref{eq:s1}) and (\ref{eq:s2}) above). 
For a diagonalization within ${\cal P}^{(q)}$ only, one obtains the 
exact leading-order term in the energy~\cite{jackson2001}. 
The contribution of all other Fock states that are elements 
of ${\cal Q}^{(q)}$ lowers the energy 
to subleading order in $N$.  (As an example, in the truncated space
$m=(0,2,4)$, diagrams which contribute to subleading order in $N$ are 
shown in Fig.~\ref{fig:4}(a) 
where the contribution of the states with $m=1$ and $m=3$ 
may be considered perturbatively).
The sum of amplitudes of all Fock states that are in 
${\cal P}^{(q)}$ is plotted in the lower panel of Fig.\,1
for those states that are downward cusps in the yrast
line for $q=2, 3$ and $4$. The amplitude sums practically vanish around the 
transitions between different values of $q$ 
where the exact yrast states become very mixed, i.e., 
there is a superposition of very
many Fock states with comparable and small amplitudes. 

Remarkably, around angular momenta where vortex states with a 
given symmetry occur as ground states, the subspace ${\cal P}^{(q)}$ 
adds up to only a fraction of the total weight of the exact 
quantum state. For the unit vortex with $q=1$, as discussed above, 
there is a single term $|0,1^N,0,0,\dots \rangle $ that has exactly $50\%$ 
of the total amplitude. 
For $q=2$, only about $30\% $ of the total amplitude 
is within ${\cal P}^{(2)}$, while the contribution of the majority of 
Fock states that belong to ${\cal Q}^{(2)}$ amounts to the remaining 70\%.
For $q=3$, the weight of the restricted subspace ${\cal P}^{(3)}$ 
decreases to about $20\%$, and for $q=4$ we obtain only about $10\%$ 
in ${\cal P}^{(4)}$. 

The ratio between the weights of  ${\cal P}^{(q)}$ and ${\cal Q}^{(q)}$ 
does not appear to be a finite-$N$ effect, but systematically persists 
for larger system sizes \footnote{We varied the particle 
number from $N=20$ which is sufficiently small to be treated with
no truncation, up to about $N=100$, where we were limited to only 
$m \le 7$ orbitals. In this range of particle numbers, we found 
that the sum of the amplitudes in ${\cal P}^{(2)}$ shows only a very 
small decrease of at most one percent, a value close to the limit 
of accuracy caused by the unavoidable truncation of the single-particle 
basis states in the case of large $N$ and $L$.}.
This becomes particularly clear when studying the Fock state amplitudes 
in the different subspaces. 
For the two-vortex state at $L/N=1.8$ for  
$N=20$, $40,  60$ and $100$ particles~\footnote{For $N\le 40$, we used 
a truncation of $m\le 14$, for $N\le 70$, $m\le 9$, 
and for $N=100$, $m\le 7$ single-particle orbitals.}, Fig.~\ref{fig:4}(b) 
shows the absolute values of the amplitudes (ordered after their 
absolute size) that are found to scale with
the particle number as $N^{1/4}$, as a function of the Fock space 
index $j$, that scales as $1/N^{1/2}$, for the subspaces 
${\cal P}^{(2)}$ (which is relatively small in dimension)
and ${\cal Q}^{(2)}$ (which is huge, containing very many states with small 
amplitudes). Fig.~\ref{fig:4}(c) shows the corresponding sums of the 
squared amplitudes $\sum _{j=1}^{j_{max}} |C_j|^2$ (here only for $N=100$) 
in ${\cal P}^{(2)}$ and ${\cal F} = {\cal P}^{(2)}\cup {\cal Q}^{(2)}$.  
The sum in ${\cal F}$ quickly saturates to unity, while in the 
restricted space ${\cal P}^{(2)}$  it saturates to only about 30\% of the total weight of the quantum state. 
We see that for particle numbers $N\ge 40$, 
the scaling in $N$ becomes nearly perfect, 
indicating that the distribution of states between ${\cal P}$ 
and ${\cal Q}$ would not change when going to even larger particle numbers. 


\section{Conclusion}

To conclude, using the method of numerical diagonalization for a few dozens 
of atoms rotating in a harmonic trap, we found that 
quasiperiodic oscillations along the yrast line 
originate from the finiteness of the system, and  
disappear in the mean-field limit of large $N$.
Furthermore, comparing the yrast state in the restricted subspace 
corresponding to the mean-field solution,   
with the exact yrast state in the full space, 
we found that it accounts for only a fraction 
of the total weight. There is additional structure in the exact state  that
persists when the system approaches the limit of large $N$, 
even though the mean-field approximation provides the yrast energy exactly in
this limit. 
 
\section{Acknowledgements}

We thank A.D. Jackson, M. Manninen, C.J. Pethick, C. Verdozzi,  
A. Wacker, and S. {\AA}berg for discussions and useful input. 
This work was financially 
supported by the Swedish Research Council and the
Nanometer Structure Consortium at Lund University, as well as  
the research networking programme POLATOM.



%

\end{document}